\documentclass[a4paper]{iopart}

\usepackage{epsfig}

\usepackage{graphicx,latexsym,xcolor}

\usepackage{tikz}

\usepackage[english]{babel}
\expandafter\let\csname equation*\endcsname=\relax 
\expandafter\let\csname endequation*\endcsname=\relax

\usepackage{amsmath}
\usepackage{bm}
\usepackage{times,amsmath,amssymb}
\usepackage{subfigure}
\usepackage{amsfonts}
\usepackage{color}
\usepackage{braket}

\begin{document}

\title[Self-organization in many-body systems with short-range interactions]{Self-organization in many-body systems with short-range interactions: clustering, correlations and topology}
\author{Ioannis Kleftogiannis}
\address{Physics Division, National Center for Theoretical Sciences, Hsinchu 30013, Taiwan}
\ead{ph04917@yahoo.com}
\author{Ilias Amanatidis}
\address{Department of Physics, Ben-Gurion University of the Negev, Beer-Sheva 84105, Israel}

\date{\today}
\begin{abstract}
We investigate the self-organization of point-particles
with short-range interactions modeled via simple 1D and 2D Hubbard-like models.
We show how various properties emerge such as, boson-like  ordering
leading to topological structures in real space, via the clustering
of the particles at discrete energy levels, which can be analyzed using a network/graph mathematical language. We calculate analytically the number of clusters,
the correlations between them and topological numbers like the Euler characteristic, deriving different organizational schemes and entanglement entropy scaling laws. All calculations are performed for an arbitrary number of particles N and energy of the system E.
Our results demonstrate how orders with diverse topological/geometric and entanglement
features, emerge in strongly interacting many-body systems,
that follow minimal microscopic interaction rules.
\end{abstract}


\section{Introduction}
Self-organization mechanisms occurring in many-body systems
that contain a large number of interacting components,
are well known to result in diverse physical phenomena.
Such phenomena include for example, the formation of exotic phases of matter and universal collective behaviors,
that are difficult to extract reductively from first-principles or perturbative approaches.
The many-body problem is also historically one of the most
difficult and computationally demanding to analyze,
due to the huge number of degrees
of freedom present, making these problems unreachable
by traditional reductionism methods applied 
to other fields of physics.
Historically one of the first self-organization 
mechanisms discovered and sufficiently understood is the long-range ordering
mechanism responsible for the formation of familiar phases of matter such as the solid,liquid and vapor phases. These phases can be described by Landau's theory based on symmetry breaking mechanisms and the renormalization group(RG) scheme developed by Leo Kadanoff and Kenneth Wilson and expanded by others\cite{landau,kadanoff,wilson}. Other more recent examples of many-body self-organization phenomena with experimentally measurable consequences are topologically ordered phases of matter. These phases lack long-range order, but are related to topological properties of the system and quantum correlations(entanglement), instead of symmetries\cite{Berezinskii,Berezinskii1,kt,kitaev1,spinhaldane,haldane0,Levin,Gu,kitaev2,kitaev3,Li,AKLT,haldane_geometry,amico,horodecki,Popkov,Hamma,wang2,Calabrese,Pollmann}.
Topologically ordered phases manifest in systems like superfluid films in the classical case\cite{Berezinskii,Berezinskii1,kt} or spin-liquids, the fractional-quantum-Hall state (FQHE)\cite{Tsui,Laughlin,Gu}
,Bose-Einstein condensates and cold-atomic systems\cite{Alba,hen,mathias,Ollikainen,hur,orth}, in the quantum case.
Apart from topology, geometry has been shown to play an important role in the physical properties of many-body states, such as those formed in the fractional-quantum-Hall-effect (FQHE)\cite{haldane_geometry}.
Other celebrated examples of self-organization mechanisms
are those encountered in non-linear many-body systems, like the well known example of self-organized criticality \cite{bak, bak1}.

Usually self-organization phenomena have a universal character, meaning that they
are independent from the microscopic details of the physical system.
Therefore in order to analyze such self-organization mechanisms one can 
either consider general arguments that can be applied to any system
independently of its microscopic details (based on symmetries for example) or choose the simplest
model with the most minimal microscopic rules (toy model).
In our paper we follow the second approach, 
examining how intricate structures emerge in 
self-organizing toy model systems, with many particles interacting
via the most minimal microscopic rules. As a paradigm we 
consider 1D and 2D Hubbard-like models
with nearest-neighbor interactions\cite{1d,2d,prefactor}.
By doing so, we show how various clustering mechanisms emerge
at discrete energy levels, resulting in formation
of particle structures with diverse geometrical
properties related to topology. We analyze the
formation mechanisms and the organizational schemes of the clusters along with the relevant topologies emerging for 1D and 2D systems, using a network/graph mathematical formalism
based on topological numbers, like the Euler characteristic. In addition we calculate analytically the correlations between the clusters in the 1D system, by using the bipartite entanglement entropy formalism, and show various entanglement behaviors appearing at different energies.

\section{1D System}
\subsection{Clustering}

\begin{figure}
\begin{center}
\includegraphics[width=0.9\columnwidth,clip=true]{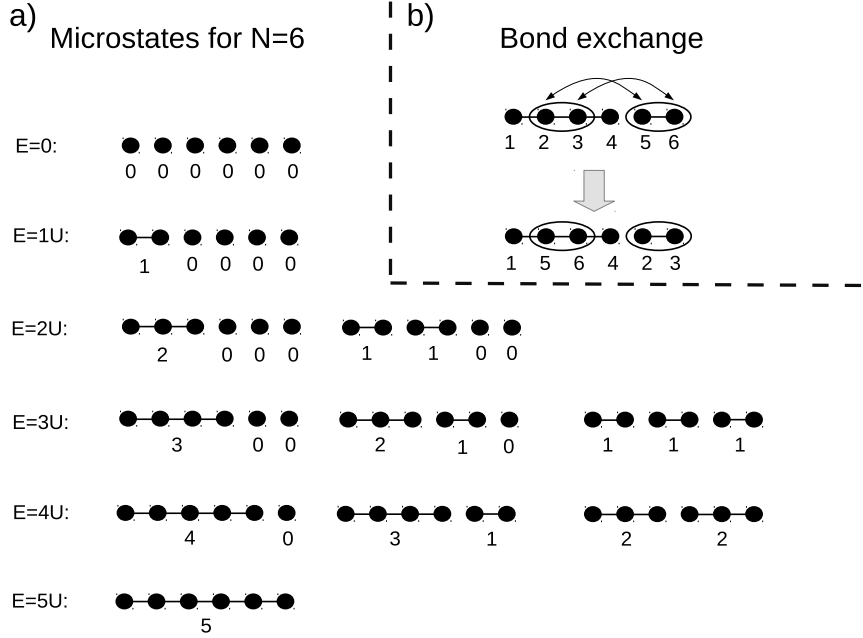}
\end{center}
\caption{a) The possible microstates (Fock states) for a Hubbard chain with nearest neighbor interactions and N=6 particles(filled circles). The empty sites in the Hubbard chain are not shown for simplicity. As the energy E increases bonds between the particles are formed, leading to clustering at discrete energy levels. Each energy level corresponds to microstates containing a fixed number of clusters, whose length is shown as a number below each cluster. The permutations of these numbers give all the possible microstates at each energy level.  b) An exchange between two bonds in a microstate that contains clusters, requires an even exchange between the original particles. Therefore the wavefunction describing the bonds, if they treated as quasiparticles, remains symmetric under their exchange, leading to bosonic statistics. }
\label{fig1}
\end{figure}
In this section we describe how a bosonic fluid with controllable properties, emerges from an one-dimensional(1D) Hubbard chain of strongly interacting particles, where only one particle is allowed per site.
An example is shown in figure \ref{fig1} for N=6 particles with a nearest-neighbor interaction U in a chain of M sites, described by the Hubbard-like Hamiltonian term\cite{1d,2d,prefactor}
\begin{equation}
H_U = U\sum_{i=1}^{M-1} n_{i}n_{i+1}.
\label{eq_h_1d}
\end{equation}
where $n_{i}=c_{i}^{\dagger}c_{i}$ is the number operator and $c^{\dagger}_{i},c_{i}$ the creation and annihilation operators for a particle at site i. For simplicity in figure \ref{fig1} we have drawn only
the particles, ignoring the empty sites in the system.
Firstly we notice that the particles self-organize at different energies condensing into clusters, resulting in the formation of various microstates at each energy with large degeneracy.
Minimal Hamiltonians like Eq. \ref{eq_h_1d} are useful to reproduce the behavior of a many-body system at the strong interaction limit when the hopping between the particles can be treated as a small perturbation which breaks the degeneracy at each energy. Although the Hamiltonian Eq. \ref{eq_h_1d} is trivially diagonal in the Fock state basis, non-trivial features emerge as the system lies in a superposition of Fock states with rich microstructures due to the short-range interaction.
The number of clusters at each energy remains constant,
for example there are always four clusters at energy E=2U. 
Every pair of neighboring particles(bond) contributes energy U in the total energy of each microstate. The bonds between two particles act as bosons. As the particles condense, the formed clusters can be considered effectively as states that can be filled by these bosons.
The number of available bosons is simply the energy of the system in units of U. The occupation number of each effective state is given by the length of the corresponding cluster. The number of available states for filling can be found, by stacking all particles together, forming a long cluster.
This cluster plus the free remaining particles
gives the total number of available effective states to be filled with bosons at energy E,
\begin{equation}
C^{1D}(N,E)=N-E.
\label{number_cluster_1d}
\end{equation}
By using Boson-Einstein statistics we can calculate the number
of microstates at each energy, that is, the degeneracy at energy E.
If we ignore the spatial gaps between the clusters, formed by the empty sites in the Hubbard chain, then we have E bosons filling $C^{1D}$ sites giving the degeneracy
\begin{equation}
D^{1D}_{P}(N,E)=\binom{C^{1D}+E-1}{E}=\binom{N-1}{E}.
\label{number_particlecluster_1d}
\end{equation}
If we ignore the length of the clusters,
(considering them as particles) then the
different ways to distribute them among M sites is\cite{1d,prefactor}
\begin{equation}
D^{1D}_{H}(M,N,E)=\binom{M-N+1}{N-E}.
\label{number_holecluster}
\end{equation}
Then in order to find the total number of microstates
at each energy we simply multiply Eq. \ref{number_particlecluster_1d}
with Eq. \ref{number_holecluster} giving the total degeneracy of the system at energy E
\begin{equation}
D^{1D}(M,N,E)=\binom{N-1}{E} \binom{M-N+1}{N-E}.
\label{degeneracy_1d}
\end{equation}

The system described above is an emergent bosonic fluid with E free bosons
represented by the corresponding bonds between the original particles. 
The E free bosons distribute among $C^{1D}$ clusters/sites. Each cluster
contains $N_i$ particles and $E_i=N_i-1$ bonds. If the total mass of the system is $N$ and the mass of each cluster is $N_i$, then we can define  an effective mass for each bond (quasi-particle) in a cluster, which would be $m_i=\frac{N_i}{(N_i-1)}>1$, satisfying $\sum_{i=1}^{C^{1D}} (N_i-1)m_{i}=N$.

The wavefunction of the system at energy E will be a superposition
of all the allowed microstates. In the effective boson picture, by ignoring
the empty space between the clusters, we will have a wavefunction of the form
\begin{equation}
\ket{\Psi} = \frac{1}{\sqrt{D_{P}^{1D}(N,E)}} (\ket{E 0 0 0 \ldots}+ \ket{(E-1) 1 0 0 \ldots}+ \ldots),
\label{Psi}
\end{equation}
where each number inside the kets represents the occupation number
(number of bosons) or length of each effective cluster/state. 
An equal superposition as in Eq. \ref{Psi} is the simplest solution that treats all
the Fock states uniformly. Other superpositions can be considered also
but are beyond the scope of the current manuscript. We expect that
the distribution of the amplitudes does not qualitatively affect the 
type of correlations emerging\cite{1d}.
Exchanging two bosons in Eq. \ref{Psi} will always lead to
an even number of particle permutations in the original wavefunction of the system as shown in figure \ref{fig1}. Therefore the wavefunction of the emergent particles(bonds) will stay symmetric independently of the symmetry/antisymmetry under exchange of the wavefunction of the original particles. Taking account of the fact that we allow only one particle per site in the Hubbard chain Eq. \ref{eq_h_1d}, the original particles can be either hard-core bosons or fermions.

In overall, our analysis shows that an 1D quantum fluid of N  fermions/hard-core bosons at energy E with short-range interactions, is equivalent to an 1D bosonic fluid with E free bosons distributed among $C^{1D}=N-E$ sites.
Also it is useful to note that Eq.1 can be mapped to an Ising model of spins 
by the substitution  $n_{i} = (\sigma_{i}^{z}+1)/2$, where
$\sigma_{i}^{z}=\pm1$. This will result in  $H_{U}=\sum_{i}n_{i}n_{i+1}=U\sum_{i}(\sigma_{i}^{z}+1)(\sigma_{i+1}^{z}+1)/4 = (U/4)\sum_{i}(\sigma_{i}^{z}\sigma_{i+1}^{z}+\sigma_{i}^{z}+\sigma_{i+1}^{z} +1)$, i.e. an Ising model with a nearest neighbor interaction term but also on-site magnetic field terms. Note that with open boundaries the magnetic fields on the edges of the chain are half the value in the bulk.

\subsection{Correlations}

\begin{figure}
\begin{center}
\includegraphics[width=0.9\columnwidth,clip=true]{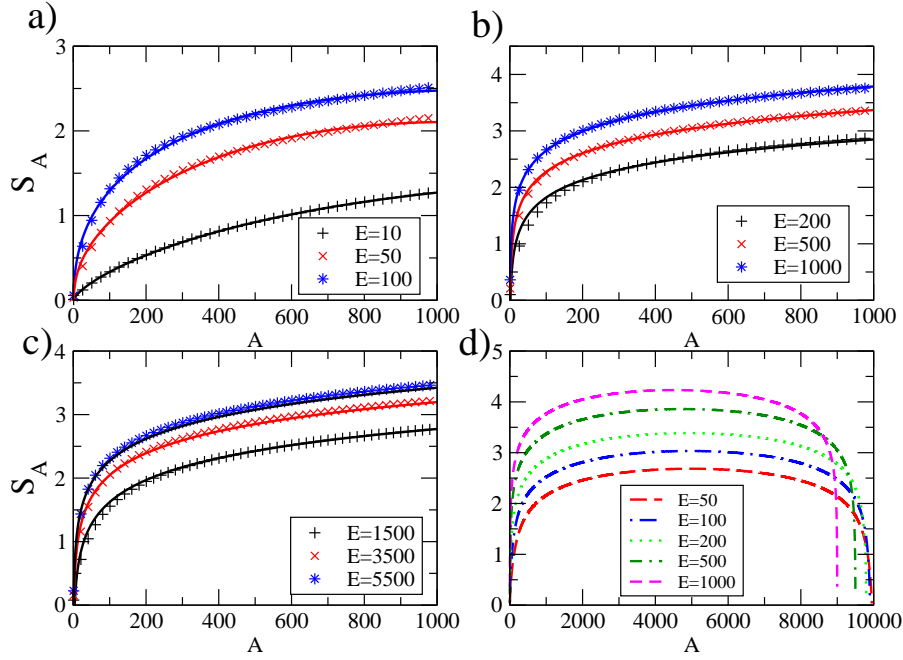}
\end{center}
\caption{The entanglement entropy $S_A$ for a partition of the 1D system containing A clusters. a)The scaling with A at low energies (E=10,50,150) ($N=10^4$) follows the form $S_{A}=a A + b A^c + d \log(A)$ with four fitting parameters $a,b,c,d$. b) The scaling of $S_A$ at higher energy (E=200, 500, 1000)($N=10^4$) has the logarithmic form $S_{A}=a \ln(A)+b$ with two fitting parameters $a,b$. Parameter $a$ reaches the asymptotic value $a=0.493709 \approx 1/2$ at $E=1000$. c) At the thermodynamic limit $N \rightarrow \infty$ and for high energies $E>1000$, the entropy is described by the analytical form Eq. \ref{eq_4_anal}-\ref{eq_5_anal} $S(N,E,A) = -\frac{1}{2} \log(A) + Q(N,E)$ shown as continuous curves. The points are plotted using the exact results calculated from Eq. \ref{eigenvalue_dm} and Eq. \ref{entropy_1d} for $N=10^5$ and agree well with the analytical result. d) The entropy for large ranges of A. The entanglement becomes stronger as E increases.}
\label{fig2}
\end{figure}

In order to quantify the correlations at a given energy we can
use the partition method that is widely applied
to calculate the entanglement in quantum many-body systems.
We apply this method to calculate the correlations
that arise when the system lies in a superposition of the possible microstates at energy E in the form of Eq. \ref{Psi}. We split the system in two partitions, one containing
A clusters and the other one the remaining $C^{1D}-A$ clusters. 
Then we can calculate the reduced density matrix of
the partition containing A clusters by tracing out the
rest of the system $\rho_{A}\equiv tr_{B}|\Psi\rangle \langle\Psi|$
, where $\ket{\Psi}$ is Eq. \ref{Psi}.
The elements of the reduced density matrix can be calculated via $\rho_{A}^{ij}=\sum_{k \in B} \Psi_{ik}^{*} \Psi_{jk}$, where $\Psi_{ik}=\frac{1}{\sqrt{D^{1D}_P}}$ is the amplitude
for each partitioned state $|ik \rangle$, where i(k) is the corresponding microstate inside each partition. We notice
that the density matrix elements are zero unless the microstates i and j contain the same number of bonds connecting neighboring particles.
Therefore, the density matrix splits in blocks each one corresponding
to a fixed number of bonds m inside the partition A. Each block contains $D^{1D}_{P}(A,m)$ elements coming from all the possible ways to arrange m bonds in A clusters as given by Eq. \ref{number_particlecluster_1d}. In addition, each block is a
full matrix with identical elements $D^{1D}_{P}(C^{1D}-A,E-m)/D^{1D}_{P}(C^{1D},E)$.
Consequently, each block in the density matrix gives only one eigenvalue 
\begin{equation}
f(C^{1D},E,A,m)=D^{1D}_{P}(A,m) \frac{D^{1D}_{P}(C^{1D}-A,E-m)}{D^{1D}_{P}(C^{1D},E)}.
\label{eigenvalue_dm}
\end{equation}
Note that the fraction in the above equation, for $A=1$,
gives the probability of cluster containing m bonds, which is essentially the probability distribution of the length of the clusters
\begin{equation}
P(C^{1D},E,m)=\frac{D^{1D}_{P}(C^{1D}-1,E-m)}{D^{1D}_{P}(C^{1D},E)}.
\label{cluster_propability_distribution}
\end{equation}
The von Neumann entanglement entropy $S\equiv -tr(\rho_{A}\log(\rho_{A}))$
of the partition A can be calculated, by summing over
the eigenvalues of each block of the reduced density matrix
\begin{equation}
S(C^{1D},E,A)\equiv -\sum_{m=0}^{E} f(C^{1D},E,A,m) \log f(C^{1D},E,A,m).
\label{entropy_1d}
\end{equation}
Using this method we have calculated the bipartite entropy
at different energies as a function of the partition size A
for $N=10^4$. The results are shown in figure \ref{fig2}.
For low energies near the ground state(E=0), shown in figure \ref{fig2}a, all the points can be fitted
by an expression $S^{1D}_{A}=a A + b A^c + d \log(A)$ with
four fitting parameters $a,b,c,d$ represented by the continuous curves. Therefore, at low
energies the entanglement entropy of the 1D system follows a mixture of power law and logarithmic scaling with the partition size. Despite its low energy the system does not obey the area law, $S \sim L^{D-1} \Rightarrow S \sim constant$, expected for the ground state of gapped many-body systems. The logarithmic term however is encountered also for the critical phases in the ground state of Ising spin chains\cite{kitaev1,calabrese1}. 

On the other hand, we have obtained a pure logarithmic
scaling for high energies $(E>100)$ shown in figure \ref{fig2}b.
All points at this energy range can be fitted sufficiently well by the  expression $S^{1D}_{A}=a \log(A)+b$ using two fitting parameters $a,b$. Parameter $a$ reaches gradually the asymptotic value $a=0.493709 \approx 1/2$ as the energy $E=1000$ is reached.
A similar universal prefactor that takes values 1/3 for free bosons and 1/6 for free fermions has been shown to be related to conformal field theories effectively describing these systems\cite{kitaev1,calabrese1}.

\subsubsection{Thermodynamic limit $N \rightarrow \infty$}
In this subsection we derive analytically the scaling
behavior of the entropy with partition size for
a large number of particles ($N \rightarrow \infty$).
By setting $\alpha= \frac{N-E}{A}$ Eq. \ref{eigenvalue_dm} becomes 
\begin{equation}
\begin{aligned}
f^{1D}(N,E,\alpha,m)=D^{1D}_{P}\left( \frac{N-E}{\alpha},m \right) \\
\frac{ D^{1D}_{P}( (N-E)(1 - \frac{1}{\alpha }),E-m)}{D^{1D}_{P}(N-E,E)}.
\end{aligned}
\end{equation}
Then we can take its thermodynamic limit ($N \rightarrow \infty$) obtaining,
\begin{equation}
\begin{aligned}
  \lim_{N\to\infty} f^{1D}(E,\alpha,m)=
   (\alpha - 1)^{(E-m)} \alpha^{-E } \binom{E}{m}.
  \end{aligned}
  \label{eq1_anal}
\end{equation}
Then, by setting $p=\frac{1}{\alpha}$ and $q=1-p$ Eq. \ref{eq1_anal}
becomes,
\begin{equation}
\begin{aligned}
f^{1D}(E,p,q,m)=
    p^{m } q^{(E-m)}\binom{E}{m}.
  \end{aligned}
  \label{eq2_anal}
\end{equation}
In the  limit $Epq\gg 1$  we obtain the normal(Gaussian) distribution with mean value $\mu$ and variance $\sigma$,
\begin{equation}
f^{1D}(E,\alpha,m) \approx \frac{1}{\sqrt{2\pi \sigma^{2}}} e^{- \frac{(m-\mu)^2}{2\sigma^{2}}}
\label{eig_gaussian}
\end{equation}

\begin{align}
 \mu &= E/\alpha
 \label{eig1}
 \\
 \sigma &= \sqrt{\frac{E}{\alpha}\left( 1- \frac{1}{\alpha}\right)},
 \label{eig2}
\end{align}
which satisfies,
\begin{align}
 \int_{0}^{E} f^{1D}(E,\alpha,m) dm =1.
\end{align}
The  von Neumann entropy of partition A can be calculated
by transforming the summation in Eq. \ref{eigenvalue_dm} to an integral as,
\begin{align}
S(E, \alpha) &=-\sum_m  f^{1D}(E,\alpha,m) \log f^{1D}(E,\alpha,m)  \approx\\
 &-\int_0^{E} f^{1D}(E,\alpha,m) \log f^{1D}(E,\alpha,m) dm.
\end{align}
After some algebra, in the limit of $E\gg1$, we obtain
\begin{equation}
\begin{aligned}
 S(E,\alpha) = -\frac{1}{2} \left( Log\left[ 2E\left(1 -\frac{1}{\alpha} \right) \frac{1}{\alpha} \right] + Log[\pi] \right ) \\
  -\frac{1}{4} \left(2 -\frac{ \frac{E}{\alpha}  }{\sqrt{E \left(1 -\frac{1}{\alpha} \right) \frac{1}{\alpha}}} e^{-
             \frac{ \left(  \frac{E}{\alpha} \right)^{2} }{2E\left(1 -\frac{1}{\alpha} \right) \frac{1}{\alpha}}}  \sqrt{\frac{2}{\pi}} \right)
\end{aligned}
\label{eq_4}
\end{equation}
where $\alpha > 1$.
Expanding the above equation, taking also account of  $N \gg E$, $\alpha= \frac{N-E}{A} \approx \frac{N}{A}$ and $N \gg A$ we get 
\begin{equation}
\begin{aligned}
 S(N,E,A) = -\frac{1}{2} \log(A) + Q(N,E),
\end{aligned}
\label{eq_4_anal}
\end{equation}
with
\begin{equation}
\begin{aligned}
Q(N,E)=\frac{1}{2} (\log(N) - \log( \pi) - \log(2E) - 1).
\end{aligned}
\label{eq_5_anal}
\end{equation}
A comparison between the analytical result Eq. \ref{eq_4_anal}-\ref{eq_5_anal} and
the exact result using Eq. \ref{eigenvalue_dm} and Eq. \ref{entropy_1d} is shown in figure \ref{fig2}c. 
In addition from Eq. \ref{eq_5_anal} we can see
that the entanglement grows logarithmically, becoming stronger, as the energy of the system E is increased.
This behavior can be seen in figure \ref{fig2}d.

The prefactor $1/2$ in front of the log in Eq. \ref{eq_4_anal}
can be understood mathematically as a consequence of the Gaussian form in Eq. \ref{eig_gaussian}. Physically this prefactor can be related to certain permutational symmetries present in the wavefunction of the system, which is expressed as a superposition of the different microstates allowed at 
energy E(Eq. \ref{Psi}). Exchanging any two cluster
lengths or alternatively any two occupation numbers in the effective boson picture, leaves the wavefunction unchanged, i.e, the wavefunction is permutationally invariant under cluster length permutations.
The prefactor $1/2$ is also present 
in the logarithmic growth of the entropy for the ground state of our system (E=0), which contains microstates with single particles as clusters and at least one empty site between them\cite{prefactor}.
The prefactor $1/2$ for this ground state is also due to partial permutational invariance of the respective wavefunction.

\subsection{Topology}
The number of clusters contained in each microstate remains constant at a fixed energy. Therefore, the number given by Eq. \ref{number_cluster_1d} can be used as a topological number to characterize the microstates with the same energy. Loosely speaking the clusters or the spatial gaps between them could be considered as holes in an 1D manifold. Alternatively, each collection of microstates at a fixed energy can be thought as a set of vertex lines whose total length, determined by the energy, remains constant. Each vertex line
is a path graph $P_n$ where $n$ is the number of its vertices.

In mathematical terms the microstates at each energy form a set of cofinite topology that has a topological invariant, the number of clusters/line segments. Any line cluster with a finite number of vertices, can be transformed to any other by the homeomorphic processes of subdivision (adding vertices) or smoothing out vertices, which is essentially like increasing or reducing the length of the clusters. Therefore all the microstates at the same energy can be continuously deformed between each other. On the other hand changes between microstates with a different number of clusters, which lie at different energies, require  non-homeomorphic processes such us splitting a line segment in two. In conclusion the
microstates at different energies have different topologies.

\section{2D system}
\subsection{Clustering}

\begin{figure}
\begin{center}
\includegraphics[width=0.9\columnwidth,clip=true]{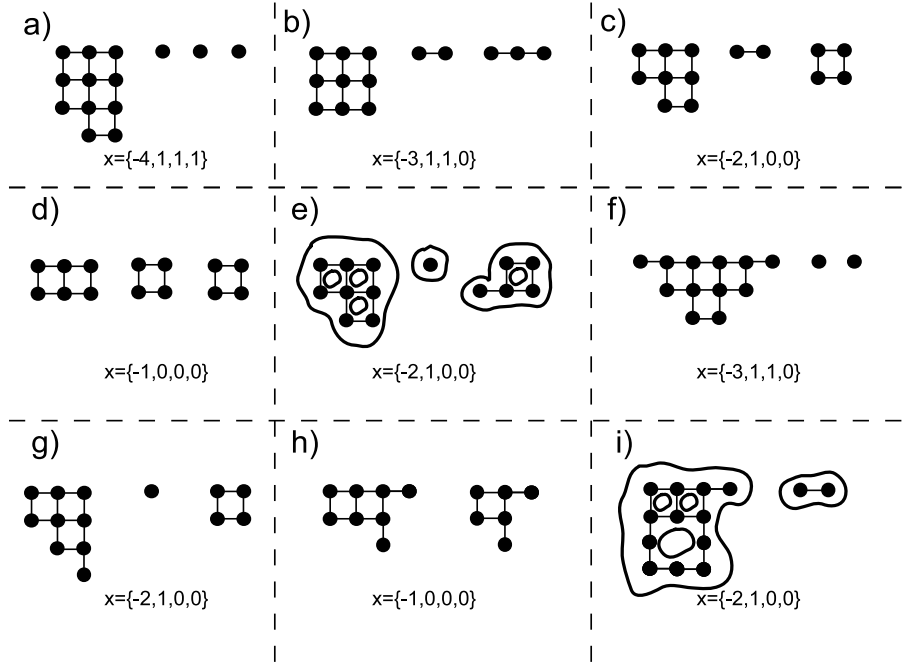}
\end{center}
\caption{Some of the possible microstates for a 2D Hubbard square lattice with nearest neighbor interactions, with N=14 particles(filled circles) at energy E=15. The clustering leads to networks/graphs, consisting of purely vertex-line clusters and clusters with cycles (closed vertex-lines) which are not possible in the 1D system. The microstates can be described by a set of numbers. Each number represents the Euler number for each individual cluster in the described microstate. Microstates described by the same set of Euler numbers have the same topology and can be deformed continuously between each other, for example microstates e) and i).The encircled areas denote the 2D manifolds that can be deformed.}
\label{fig3}
\end{figure}

In 2D the particles form more intricate structures than in 1D,
due to the additional degrees of freedom present.
The system can be described by the following Hubbard Hamiltonian\cite{2d}
\begin{equation}
H = U \sum_{x=1}^{M_{x}}\sum_{y=1}^{M_{y}}(n_{x,y}n_{x+1,y} +n_{x,y}n_{x,y+1}) 
\label{h_2d}
\end{equation}
where $n_{x,y}=c_{x,y}^{\dagger} c_{x,y}$ is the number operator with $c_{x,y}^{\dagger},c_{x,y}$ being the creation and annihilation operators for spinless particles at site with coordinates x,y, in a square lattice with periodic boundary conditions $N+1 \rightarrow 1$. 

The self-organization of the interacting particles can be split in two clustering mechanisms. One
is the same as in 1D, resulting in line clusters whose total length is determined by the energy of the system. The other mechanism is related to formation of closed vertex-lines(cycles) that is not possible in 1D. One way to help categorize all the possible structures generated by these two mechanisms, is by using the Euler characteristic $\chi$ of the graph/network structures formed by the particles as they condense. The Euler characteristic of a graph can be defined as (handshake definition)
\begin{equation}
\chi=N-\textit{E},
\label{eq_euler}
\end{equation}
where N is the number of vertices and $\textit{E}$ the number of edges between the vertices in the graph.
Each cluster will have its own number of vertices(particles) $N_i$,edges(bonds between particles) $E_i$ and an Euler number $\chi_i=N_i-E_i$.
In the graph mathematical language the structures in 2D
consist of either path graphs $P_n$, grid graphs(square lattices) $P_n \times P_m$ with $N_i=nm$ or compositions/combinations of both\cite{epstein}. 

If there are $C^{2D}$ clusters in total at energy E,
then the following expression should be satisfied
\begin{equation}
\chi=\sum^{C^{2D}}_{i=1} \chi_i.
\label{eq_euler_sum}
\end{equation}
As for the 1D system, the number of clusters, along with the allowed values of $\chi_i$, are both determined by the energy of the system and the total number of particles N. Each microstate can be described by a set of $2C^{2D}$ numbers, either, $\{N_1,...,N_{C^{2D}},E_1,...,E_{C^{2D}}\}$, $\{\chi_1,...,\chi_{C^{2D}},N_1,...,N_{C^{2D}}\}$ or $\{\chi_1,...,\chi_{C^{2D}},E_1,...,E_{C^{2D}}\}$. For our analysis we choose
the set $\{\chi_1,...,\chi_{C^{2D}},E_1,...,E_{C^{2D}}\}$.
In the network/graph mathematical language, the problem consists essentially of investigating the different ways E edges can be distributed among N vertices and the resulting topological structures emerging, represented by $C^{2D}$ clusters. In 1D the only possible values of $\chi_i$ are $0$ when no cluster is present and $1$ for the clusters that consist of either a single particle or a line of particles (vertex-line). In 2D however additional negative values of $\chi_i$ are allowed since the clusters can contain closed vertex-lines (cycles), which result in $N_i \le E_i$. In general
in both 1D and 2D, the Euler characteristic will take values in the interval
\begin{equation}
\chi_i \in [\chi_{min},1].
\label{eq_euler_limits}
\end{equation}
Some examples of the two clustering mechanisms for 2D can be seen in figure \ref{fig3} for various microstates in a system with $N=14,E=15$. We notice that microstates with the same energy, can contain a different number of clusters in them, unlike in 1D. For convenience, in our analysis we consider a fixed number of clusters $C^{2D}$
at each energy, which counts the real clusters but also the empty clusters that have $(\chi_i=E_i=0)$. The number of vertices $N_i$ in each cluster constrains the corresponding number of edges $E_i$ that can fit in the cluster, by the following equation
\begin{equation}
 E_i \le 2(N_i - \sqrt{N_i}). 
\label{edges_constraint}
\end{equation}
The minimum value $\chi_{min}$ for all microstates at energy E, can be obtained by stacking all the E edges/bonds in one of the clusters.
The number of vertices in this cluster can be found 
by using the equality in Eq. \ref{edges_constraint} with $E_i=E$,
\begin{equation}
N_1=\frac{1}{2} (1+E+\sqrt{1+2E}).
\label{vertices_cluster}
\end{equation}
In case that the above equation does not give an integer value, then $N_1$ has to be rounded to the next largest integer.
The cluster containing $N_1$ vertices gives the lowest possible value of the Euler characteristic at energy E,
\begin{equation}
 \chi_{min}(N,E)=N_1-E=\frac{1}{2} (1-E+\sqrt{1+2E}). 
\label{eq_euler_min}
\end{equation}
This cluster plus the remaining free particles(vertices) gives the total number of clusters $C^{2D}(E,N)$ available for filling with edges, in analogy to the idea applied in the 1D system,
\begin{equation}
C^{2D}(N,E)=N-N_1+1=N-\frac{1}{2} (-1+E+\sqrt{1+2E}).
\label{cluster_number_2d}
\end{equation}
This is essentially the maximum number of clusters for each microstate at energy E. Notice that the corresponding number of vertices in 1D, for the cluster that contains all the edges is $N_1=E+1$,
which is replaced by Eq. \ref{vertices_cluster} for the 2D case.
Alternatively the number of clusters can be found
by $C^{2D}=N-N_1+1=N-(\chi_{min}+E)+1=\chi-\chi_{min}+1$.

In principle, by using Eq. \ref{eq_euler}-\ref{cluster_number_2d} we can find all the possible cluster structures in the 2D system at energy E for N particles, characterized by the set of $2C^{2D}$ numbers $\{\chi_1,...,\chi_{C^{2D}},E_1,...,E_{C^{2D}}\}$.
We analyze the procedure below.

We have $C^{2D}$ Euler numbers forming the set $\{\chi_1,...,\chi_C\}$, which partly characterizes the cluster structures, if their corresponding number of edges characterized by the set $\{E_1,...,E_{C^{2D}}\}$, is ignored. The integer values of $\chi_i$ for each cluster are constrained by Eq. \ref{eq_euler_limits} and Eq. \ref{eq_euler_min}. The number of all possible configurations of $\chi$ can be found by distributing $\left| \chi_{min} \right|$ bosons among $C^{2D}$ states, via the binomial 
\begin{equation}
D^{2D}_{x}(N,E)=\binom{C^{2D}+\left| \chi_{min} \right|-1}{\left| \chi_{min} \right|}.
\label{number_Euler_2D}
\end{equation}
In addition, each cluster structure for a fixed set $\{\chi_1,...,\chi_{C^{2D}}\}$
can contain different distributions of edges $\{E_1,...,E_{C^{2D}}\}$, that give the same total $\chi$ (Eq. \ref{eq_euler_sum}).

\subsubsection{Clusters with $\chi$ $\ne$ 0}
A cluster with Euler $\chi_{i} \ne 0$ will contain a minimum number of edges $E_i^{min}$, which can be derived by substituting $N_i=\chi_i+E_i$ in Eq. \ref{edges_constraint}, giving,
\begin{equation}
E_i^{min} = 2 (1-\chi_i+\sqrt{1-\chi_i}).
\label{Euler_edges_constraint}
\end{equation}
If the above equation does not give an integer value, then it has to be rounded to the next largest integer.
Notice that the same idea applies for the 1D system, but since the only possible structures are vertex-lines in this case, then we always have $\chi_{min}=1$ and $E_i\ge 0$. 
This means that, in 1D there are no constraints on the arrangement of the edges among the clusters.
In contrast, for the 2D system the arrangement of the edges is constrained by Eq. \ref{Euler_edges_constraint}.

\subsubsection{Clusters with $\chi$=0}
The minimum number of edges for a cluster with $\chi_i \ne 0$ is determined by Eq. \ref{Euler_edges_constraint}. Clusters with $\chi_i=0$ have to be treated separately however, since there are two possibilities for their minimum number of edges. Euler $\chi_i=0$ represents either an empty cluster with $N_i=E_i=\chi_i=0$ , or a cluster with at least one cycle (closed vertex-line) with $N_i=E_i,\chi_i=0$ giving a minimum number of edges $E^{min}_{i}=4$.

If there are n clusters with $\chi_i=0$ then there will be $2^n$ different distributions of minimum edges,
since each cluster with $\chi_i=0$ can have either zero or four minimum number of edges, as stated above. 
If there are $C^{\chi=0}_{E=0}$ empty clusters and $C^{\chi=0}_{E \ge 4}$ clusters with at least one cycle, then
the number of $\chi \ne 0$ clusters is $C_1^{2D}=C^{2D}-C^{\chi=0}_{E=0}-C^{\chi=0}_{E \ge 4}$. Then there will be
\begin{equation}
E_{free}= E-[\sum^{C_1^{2D}}_{i=1} E^{min}_i + 4 C^{\chi=0}_{E \ge 4}]
\label{edges_bosons_test}
\end{equation}
edges which can be distributed freely among $C_{2}^{2D}=C^{2D}-C^{\chi=0}_{E=0}$ clusters, which are left after excluding the empty clusters.
The minimum number of edges for the $\chi \ne 0$ clusters $E^{min}_i$ is given by Eq. \ref{Euler_edges_constraint}. The number of different ways to distribute the free edges will be given by a summation over $2^n$ binomials as,
\begin{equation}
D^{2D}(N,E)=\sum_{i=1}^{2^n} \binom{C_{2}^{2D}+E_{free}-1}{E_{free}}.
\label{number_Euler_2D_test}
\end{equation}
The sum runs over the different combinations
of minimum edges determined by the number of clusters with $\chi_i=0$. Each binomial in Eq. \ref{number_Euler_2D_test}
corresponds to a different configuration/set of minimum edges which gives a different number for $E_{free}$ and $C_{2}^{2D}$.
For a fixed number of non-empty clusters $C_{2}^{2D}$, the free edges are dangling bonds(vertices connected with one edge) that stack on top of the clusters that contain the minimum number of edges. For example the microstate in figure \ref{fig3}c has one
free dangling bond/boson which can be rearranged
as shown in figure \ref{fig3}g, conserving the same set
of Euler numbers $\{-3,1,1,0\}$.
Another remark is that when there are not enough edges
to fill all the $\chi=0$ clusters with four edges, then $E_{free} < 0$ and the respective binomial in Eq. \ref{number_Euler_2D_test} is zero.

From our analysis we can see that 
the interacting particles in 2D condense
into a fluid of E bosons distributed
among a number of clusters/states determined by N and E, but constrained also by the minimum number of edges/bosons that fit in each cluster. There are two major differences with the 1D system. Firstly, in 1D the number of clusters inside the microstates with the same energy is fixed.
Secondly, the edges/bosons in 1D distribute freely among this fixed number of clusters/states without any constraints.
Consequently, if we interpret the clusters
as composite particles, then their number in 1D
is conserved at a fixed energy, in contrast to 2D.
A fractionalization mechanism can be easily distinguished
by calculating the effective mass of the edges/bonds/quasi-particles.
This is $m_i=\frac{N_i}{E_i}$. Since in 2D the cluster structures can have $E_i>N_i$, which is not possible in 1D, each edge in the cluster has a fractionalized effective mass $m_i<1$. 

We remark that in principle, the core of the above results could be extended to any random network that can be arranged as a planar graph, one that contains no overlapping edges, formed by the interactions between many particles. In general, the total number of bonds/edges between two particles determines the energy of the system. These edges can be distributed among N particles/vertices
giving many different microstates with diverse topological/geometrical cluster structures, similar to the ones that we have demonstrated specifically for the square lattice, which is a grid graph.

\subsection{Topology}
The number of cycles(closed vertex lines) in each cluster can be used as a topological invariant to characterize the different cluster configurations,
corresponding to different sets $\{\chi_1,...,\chi_{C^{2D}}\}$. 
For instance, the cycles are five, four and three
for the cluster residing on the left side of figure \ref{fig3}a,\ref{fig3}b,\ref{fig3}c.
The number of cycles can be changed for example, by removing one bond which opens/breaks a cycle, creating dangling bonds(vertices connected with one edge). This act corresponds to a non-continuous deformation analogous to the process of pocking a hole through a sphere
to create a torus, two topologically distinct geometrical shapes.
The Euler number is analogous to the number of cycles in a cluster.
This can be understood by starting with one cycle with $\chi=0$.
Every additional cycle in the cluster will change $\chi$ by $V-E=V-(V+1)=-1$
resulting in
\begin{equation}
\chi=-cycles+1.
\label{euler_loops}
\end{equation}
The above formula is consistent with Euler's formula
for planar graphs $V-E+F=2$ where F are the faces in the graph.
For our case $F=cycles+1$ since F counts the faces enclosed by the square
lattice cluster structures(cycles), along with the external face in the surface
where the clusters are embedded. Since $\chi=V-E$ we have 
$\chi+cycles+1=2 \Rightarrow \chi=-cycles+1$.
This formula is analogous to the respective formula that connects
the Euler characteristic and the genus (number of holes) $g$ in topological objects described by differential geometry $\chi=2(-g+1)$. Geometrical
shapes with the same $g$ correspond to the same topology. 

Microstates that are described by
the same set $\{\chi_1,...,\chi_{C^{2D}}\}$ have the same topology
and the corresponding cluster structures can be continuously deformed
between each other, for example the clusters in figure \ref{fig3}e and figure \ref{fig3}i.
Note that when the cluster contains only one cycle, like
the rightmost cluster in figure \ref{fig3}e,
it can be shrunken down to an empty cluster, which is
understood topologically by taking a circle and shrinking
it's radius to zero.
Different sets $\{\chi_1,...,\chi_{C^{2D}}\}$ correspond
to cluster structures with different topologies.
Notice that the number of clusters inside each microstate at each energy is not constant,
unlike the 1d case, and therefore it cannot be used as a topological invariant. In overall, we see that while in the 1D system each sub-Hilbert space of microstates at energy E defines a topology characterized by the number of clusters, in 2D each sub-Hilbert space at energy E contains microstates with different topologies, characterized by the set of Euler numbers describing the corresponding cluster structures.

We remark that Eq. \ref{euler_loops} is valid
for any planar graph, one with no overlapping edges, even random graphs formed
by interactions between many particles.
In this sense the topological properties described above
are valid for any many-body network that can be
arranged as a planar graph.

\section{Summary and conclusions}
We have investigated the self-organization of point-particles 
with strong short-range interactions, modeled
by simple Hubbard-like models. We have demonstrated
topological clustering mechanisms and boson-like
ordering of the particles for 1D and 2D models.

In 1D the interacting particles condense into line clusters
of variable length, whose number remains constant at a fixed energy, acting as a topological invariant characterizing all the corresponding microstates.
In this sense, the interacting particles in 1D, form cofinite topological sets of line segments at different energies.
In 2D richer particle structures emerge due to the formation of closed vertex-lines(cycles) in the clusters. The respective microstates at a fixed energy,
can be categorized according to the different sets of Euler numbers, which describe the graph/network structures formed by the clusters. The number of clusters contained in each microstate at a fixed energy,  is not a constant, unlike in 1D. Therefore, the number of composite particle structures, formed by the clusters in 2D, is not conserved at an energy, in contrast to 1D. 

For both the 1D and 2D cases, we have found that the bonds between the particles act as free bosons filling a number of states determined by the total number of particles in the system and its energy, giving rise to bosonic fluids, with controllable properties. In 2D, constraints in the filling rules of the states arise, as a result of richer organizational schemes compared to 1D, due to the cycles in the clusters. 

For the 1D system we have calculated the correlations
between the clusters using the bipartite entanglement
entropy formalism. For high energies a logarithmic scaling of the entropy with the partition size can be observed, as in critical quantum systems, with a universal prefactor 1/2 related to permutational symmetries of the wavefunction describing the system. On the other hand, a mixture of power-law and logarithmic scaling behavior occurs at lower energies near the ground state of the system. 

In overall, we have demonstrated simple self-organization
mechanisms in many-body systems with short-range interactions, which give rise to network/graph structures in real space,
with diverse topological and more general geometrical features. Apart from their fundamental significance such self-organization mechanisms could be potentially useful in realizing or designing states with controllable geometrical features in cold-atomic systems, by tuning the strength and the range of interaction between the particles.

\section*{Acknowledgements}
We acknowledge resources and financial support provided by
the National Center for Theoretical Sciences of R.O.C. Taiwan and the Department of Physics of Ben-Gurion University of the Negev in Israel.


\section*{References}

\begin{thebibliography}{99}
\bibitem{landau} Landau Lev D 1937  Zh. Eksp. Teor. Fiz. {\bf  7}: 19-32
\bibitem{kadanoff} Kadanoff Leo P 1966 Physics Physique Fizika 2 {\bf 263}
\bibitem{wilson} Wilson K G  1975 Rev. Mod. Phys. {\bf 47} 4 773 
\bibitem{Berezinskii} Berezinskii V L 1971 Sov. Phys. JETP {\bf 32} (3): 493-500 
\bibitem{Berezinskii1} Berezinskii V L 1972 Sov. Phys. JETP {\bf 34} (3): 610-616 
\bibitem{kt}Kosterlitz J M and  Thouless D J 1973 J. Phys. C: Solid State Phys. {\bf 6} 1181 
\bibitem{kitaev1}Vidal G, Latorre J I, Rico E  and  Kitaev A  2003 Phys. Rev. Lett. {\bf 90} 227902
\bibitem{spinhaldane}Haldane F D M 1980 Phys. Rev. Lett. {\bf 45}, 1358 
\bibitem{haldane0}Haldane F D M 1983a  Phys. Lett. A {\bf 93} 464 
\bibitem{Levin}Levin M and  Wen X-G 2006  Phys. Rev. Lett. {\bf 96} 110405 
\bibitem{Gu}Chen X,  Gu Z-C  and   Wen X-G 2010  Phys. Rev. B {\bf 82} 155138
\bibitem{kitaev2}Kitaev A and  Preskill J 2006   Phys. Rev. Lett.  {\bf 96} 110404 
\bibitem{kitaev3}  Kitaev A Y 2003 Ann. Phys. {\bf 2} 303
\bibitem{Li} Li H  and  Haldane F D M 2008 Phys. Rev. Lett.  {\bf 101} 010504 
\bibitem{AKLT} Affleck I,  Kennedy T,   Lieb  E H and  Tasaki H 1987 Phys. Rev. Lett. {\bf 59} 799  
\bibitem{haldane_geometry}  Haldane F D M 2011 Phys. Rev. Lett. {\bf 107} 116801
\bibitem{amico} Amico L,  Fazio R, Osterloh A  and  Vedral V 2008  Rev. Mod. Phys.  {\bf 80} 517 
\bibitem{horodecki}Horodecki R,  Horodecki P,   Horodecki M and Horodecki K 2009 Rev. Mod. Phys. {\bf 81} 865 
\bibitem{Popkov} Popkov V and   Salerno  M 2005 Phys. Rev. A {\bf 71} 012301
\bibitem{Hamma}Hamma A, Ionicioiu  R and  Zanardi P 2005   Phys. Rev. A  {\bf 71} 022315 
\bibitem{wang2}Wang Y,  Gulden T,  Kamenev A 2016 Phys. Rev. B {\bf 95} 075401 
\bibitem{Calabrese} Calabrese P and  Lefevre A  2008  Phys. Rev. A  {\bf 78} 032329

\bibitem{Pollmann} Pollmann F,  Turner A M, Berg E and  Oshikawa  M 2010 Phys. Rev. B {\bf 81} 064439 
\bibitem{Tsui} Tsui D C,  Stormer H L  and  Gossard A C 1982  Phys. Rev. Lett. {\bf 48} 1559 
\bibitem{Laughlin} Laughlin R B 1983  Phys. Rev. Lett. {\bf  50} 1395 (1983)
\bibitem{Alba} Alba V,  Haque M and   Luchli A M 2013 Phys. Rev. Lett. {\bf 110} 260403 
\bibitem{hen} Hen I and Rigol M  2009 Phys. Rev. B {\bf 80} 134508 
\bibitem{mathias}Scheurer M S,  Rachel S and Orth P P 2015 Scientific Reports volume {\bf 5} 8386 

\bibitem{Ollikainen}Ollikainen T,  Blinova A,  Möttönen M and  Hall D S 2019 Phys. Rev. Lett. {\bf 123} 163003 

\bibitem{hur}Hur K,  Henriet L, Petrescu A, Plekhanov K,  Roux G and  Schiró M 2016 Comptes Rendus Physique 
Volume {\bf 17} Issue 8 

\bibitem{orth}Orth  P P 2013 et al  J. Phys. B: At. Mol. Opt. Phys. {\bf 46} 134004 

\bibitem{bak}Bak P, Tang C, and  Wiesenfeld K 1987 Phys. Rev. Lett. {\bf 59} 381 
\bibitem{bak1}Bak P, Tang C, and  Wiesenfeld K 1988 Phys. Rev. A {\bf 38} 364 
\bibitem{1d}Kleftogiannis I,  Amanatidis I 2020 Eur. Phys. J. B {\bf 93}, 84 
\bibitem{2d}Kleftogiannis I and  Amanatidis I 2019 Eur. Phys. J. B  92: 198 
\bibitem{prefactor}Kleftogiannis I, Amanatidis I and  Popkov V 2019 J. Stat. Mech.  063102 
\bibitem{calabrese1} Calabrese P and  Cardy J 2009   J. Phys. A: Math. Theor. {\bf 42} 504005 
\bibitem{epstein}Eppstein D 1992 Parallel recognition of series parallel graphs Inform. and Comput. {\bf 98}  pp.41-55  


\end{thebibliography}
\end{document}